\documentclass[12pt,preprint]{elsarticle}

\usepackage{lineno,hyperref}
\usepackage{color}
\usepackage{soul}

\usepackage{amsmath,amssymb}
\modulolinenumbers[5]

\journal{New Journal of Physics}

\usepackage{color}

\begin{document}

\begin{frontmatter}

\title{Hydrodynamic Density Functional Theory\\of simple dissipative fluids}

\author{Gyula I. T\'oth}
\address{Department of Mathematical Sciences, Loughborough University, LE11 3TU Loughborough, United Kingdom}
\ead{g.i.toth@lboro.ac.uk}

\begin{abstract}
In this paper, a statistical physical derivation of thermodynamically consistent fluid mechanical equations is presented for non-isothermal viscous molecular fluids. The coarse-graining process is based on (i) the adiabatic expansion of the one-particle probability density function around Local Thermodynamic Equilibrium, (ii) the assumption of decoupled particle positions and momenta, and (iii) the variational principle. It is shown that there exists a class of free energy functionals for which the conventional thermodynamic formalism can be naturally adopted for non-equilibrium scenarios, and describes entropy monotonic fluid flow in isolated systems. Furthermore, the analysis of the general continuum equations revealed the possibility of a non-local transport mode of energy in highly compressible dense fluids.
\end{abstract}


\end{frontmatter}


\section{Introduction}

In the framework of classical continuum mechanics, mass, momentum and heat transport in fluids are described by a set of partial differential equations known as the fundamental equations of fluid mechanics. The first equations describing the dynamics of an incompressible perfect fluid were published by  Euler \cite{euler1757}. The extension of the model for compressible fluids was introduced by Laplace \cite{Laplace1816}, who discovered that the propagation of sound is an adiabatic rather than an isothermal process (adiabatic here means no heat transfer between the system and its surroundings), which therefore necessitated the introduction of an energy balance equation. In the first half of the 19$^\textrm{th}$ century, Navier and Stokes developed a momentum balance equation for isothermal viscous fluids \cite{Navier1822,Stokes1845}, which was followed by the discovery of turbulence \cite{https://doi.org/10.1002/qj.49707934126}. Classical fluid mechanics became the part of canonical knowledge by the middle of the 20$^\textrm{th}$ century \cite{Serrin1959,Landau1987Fluid,hansen90a,batchelor_2000}. Although the classical model successfully describes transport phenomena in single-component gases and liquids, it cannot describe multiphase systems with diffuse interfaces: Since the driving force of the flow is the gradient of the scalar thermodynamic pressure, the pressure must be constant (homogeneous) in equilibrium, which does not apply across diffuse equilibrium interfaces. As early as in 1901, Korteweg introduced the idea of a continuous liquid/vapor interface \cite{Korteweg1901}, which he could describe by introducing a non-diagonal reversible stress tensor in the Navier-Stokes equation. Half a century later, this led to the development of the local version of the Gibbs-Duhem relation \cite{doi:10.1063/1.432687}, which revealed that the fluid flow was driven by the gradient of the chemical potential rather than the gradient of the pressure. The theory was later extended for multicomponent systems \cite{doi:10.1098/rspa.1997.0086,doi:10.1146/annurev.fluid.30.1.139,ANDERSON2000175}, thus making the phenomenological continuum theory of fluids thermodynamically consistent. Despite the immense success of the phenomenological theory in practical applications, there are still uncharted territories on the map. The most important of these might be the theoretical description of enhanced heat transport in near-critical fluids, a phenomenon efficient cooling of pressurized-water nuclear reactors is based on. The essence of the phenomenon is that in near-critical fluids, a fast transport mode of heat arises (in addition to diffusion and convection) due to the so-called ``piston effect'' \cite{PhysRevA.41.2264}. While classical fluid mechanics successfully captures the critical slowdown of transport phenomena related to diffusion and convection, it fails to describe this anomalous heat transport mode \cite{ZAPPOLI2003713,ZAPPOLI1991269,SHEN20131}, which suggests the incompleteness of the classical phenomenological theory.
 
The key to the successful extension of classical fluid mechanics for generic fluids is the development of a deeper understanding of their microscopic origin, which has been the topic of theoretical investigations for more than 70 years. Connecting classical continuum mechanics and the Hamiltonian particle dynamics became possible with the birth of Statistical Mechanics. An early attempt to derive fluid mechanical equations for simple fluids was presented by Irving and Kirkwood \cite{doi:10.1063/1.1747782}, who provided exact (but not closed) hydrodynamic equations in terms of the many-particle probability density functions (p.d.f.'s) and interaction potentials. Closed equations were developed later by Ilyushin, who used phenomenological constitutive relations to approximate the unknown terms in the equations for gases \cite{Ilyushin}. A more rigorous approach was then proposed by Chapman and Enskog, which relied on writing the solution of the Boltzmann equation as a series of the Knudsen number \cite{chapman1970mathematical}. While the first results were only available for dilute systems, the theory was later extended for dense gases \cite{Davis1973,EU1979187,EU1979230,EU1979423}. Since the Boltzmann equation can be obtained by replacing the two-particle p.d.f. and the pair potential in the first equation of the Bogoliubov–Born–Green–Kirkwood–Yvon (BBGKY) hierarchy by a collision integral, the hydrodynamic equations derived in the framework of the kinetic theory only refer to gases. The problem has been partially resolved by using the Revised Enskog Theory \cite{VANBEIJEREN1973437}, which was successfully used to develop closed hydrodynamic equations describing the freezing transition in the hard-sphere system \cite{10.1063/1.4900499}. The Enskog theory was also extended for continuous potentials, where estimations for the macroscopic transport coefficients were given \cite{10.1063/1.1355978,2001PhyA..298..101M}. 

While deriving thermodynamically consistent continuum equations from the microscopic dynamical equations for simple dissipative fluids of arbitrary density and pair potential is still an unresolved problem, such equations can be derived for colloidal suspensions, where macroscopic-size particles float in a carrier liquid. The dynamics of these systems is studied in the framework of the so-called Dynamical Density Functional Theory (DDFT)  \cite{doi:10.1080/00018737900101365,doi:10.1063/1.478705,doi:10.1080/00018732.2020.1854965}. The early versions of the theory provided a single equation of motion for over-damped Brownian systems, while inertial effects were incorporated later \cite{PhysRevLett.94.183001,PhysRevLett.101.148302,Rex2009,10.1063/1.4883520,doi:10.1063/1.4913636,10.1063/1.2188390,10.1063/1.2724823,Marini_Bettolo_Marconi_2008}. The most recent versions of the theory operate with momentum and energy-type fields in addition to the one-particle density, for which closed hydrodynamic equations are developed  \cite{PhysRevLett.109.120603,Goddard_2013,Duran-Olivencia2016,CHAVANIS20111546,Goddard2021}. For colloidal systems with small (but still moderate) friction coefficients, significant discrepancies have been found between the theory and experiments \cite{doi:10.1063/1.3054633}, which can be eliminated by applying systematic corrections \cite{10.1063/1.2188390,10.1063/1.2724823,Marini_Bettolo_Marconi_2008}. Based on the results, one might be tempted to take the zero friction - small particle mass limit of the theory to obtain hydrodynamic equations for molecular systems. Although the resulting continuity and Navier-Stokes equations look formally correct, the approach is precarious as the Langevin picture itself collapses in this limit for the following reasons: (i) As the mass of the ``colloid'' particle becomes comparable to that of the carrier fluid, the heat bath cannot be modelled as a continuum any more, and therefore one needs to return to full molecular resolution. (ii) In addition, in a Langevin system of zero friction the system is fully decoupled from the medium, and therefore the adiabatic heat bath picture breaks down. To resolve the problem, Archer started from the Hamiltonian equations of motion and developed an exact but not closed dynamical equation for dense molecular fluids \cite{Archer_2006}. The equation was made closed by using a simple approximation for the dissipative term, however, the resulting equation differs from the one emerging from the phenomenological theory for small density variations in the Stokes limit \cite{doi:10.1063/1.4913636}. The most recent advance in the field is the development of the Power Functional Theory by Schmidt and co-workers  \cite{10.1063/1.4807586,RevModPhys.94.015007} and its reconsideration by Lutsko \cite{Lutsko2021}. This approach is based on the generalisation of Onsager's variational principle, and allows the derivation of mechanical equations for dissipative fluids. Thermodynamically consistent, closed fluid mechanical equations for dense dissipative molecular fluids, however, have not yet been derived on the statistical mechanical basis.

In this paper, we use a statistical mechanical approach to derive thermodynamically consistent fluid mechanical equations for simple dissipative fluids. {First, exact coarse-grained equations will be derived on the basis of the operator representation of the Hamiltonian dynamics. To obtain practical continuum equations for slow fields, we will use three main assumptions as follows. (i) Leading-order corrections around Local Thermodynamic Equilibrium (LTE) will be used to obtain general equations for the coarse-grained density, velocity, and temperature fields for dissipative systems. (ii) To obtain a closed equation for the temperature, we assume that the expectation of the momentum of a particle w.r.t. fixed particle positions only depends on the position of the selected particle, and the average velocity is determined by the hydrodynamic velocity field. (iii) To obtain thermodynamically consistent Navier-Stokes equations, we give a non-isothermal extension of the variational formalism. We will show that the variational formalism provides a general Helmholtz free energy functional with a gauge invariant term not contributing to the dynamics. It will be shown that equilibrium in the isolated system coincides with the stationary points of the conventional entropy functional w.r.t. constant total mass and energy. By eliminating the gauge invariant term, a class of free energy functionals will be identified for which the dynamics is entropy monotonic (i.e., the entropy production rate is non-negative), thus resulting in a thermodynamically consistent continuum theory.} Finally, it will be shown that the general energy balance equation has a non-local term which is absent in the phenomenological/kinetic theory. This term may become dominant in highly compressible dense fluids, and is therefore crucial to describe enhanced heat transport in near-critical fluids.

\section{General coarse-grained equations}

Let $\mathbf{r}_i(t)$ and $\mathbf{p}_i(t)$ (where $i=1\dots N$) denote the position and momentum of particle $i$ in an isolated system of $N$ {identical} classical particles. The Hamiltonian of the system reads as $\hat{H}=\hat{K}+\hat{V}$, where $\hat{K}=\frac{1}{2}\sum_i \frac{|\mathbf{p}_i|^2}{m}$ {is the kinetic energy, and $m$ the particle mass. The potential energy reads as $\hat{V}=\sum_{n=1}^\infty \frac{1}{n!}\sum_{i_1,i_2,\dots,i_n}u_n(\mathbf{r}_{i_1},\mathbf{r}_{i_2},\dots,\mathbf{r}_{i_n})$, where $u_1(\mathbf{r}_1)$ describes an external field, while $u_n(\mathbf{r}_1,\mathbf{r}_{2},\dots,\mathbf{r}_n)$ is the many-body interaction potential for $n \geq 2$. To avoid self-interactions, $u_n(.)=0$ is required if any two of the arguments coincide. Furthermore, $u_n(.)$ is assumed to be invariant for the permutation of its arguments, and the Fourier transform of $u_n(\mathbf{r}_1,\mathbf{r}_2,\dots,\mathbf{r}_n)$ w.r.t. $\mathbf{r}_i$ must exist for $i=1,2,\dots,n$ and $n \in \mathbb{N}$.} The time evolution of the system is governed by the canonical equations {$\dot{\mathbf{r}}_i(t) = \frac{\mathbf{p}_i(t)}{m}$ and $\dot{\mathbf{p}}_i(t) = - \frac{\partial \hat{V}}{\partial \mathbf{r}_i}$}. {Using} the Dirac-delta formalism first introduced by Irving and Kirkwood  \cite{doi:10.1063/1.1747782}, the microscopic mass, momentum, and kinetic energy density are defined as:
\begin{eqnarray}
\label{Mden}&&\hat{\rho}(\mathbf{r},t) \equiv \sum_i m \, \delta[\mathbf{r}-\mathbf{r}_i(t)] \,, \\
\label{Mmom}&&\hat{\mathbf{g}}(\mathbf{r},t) \equiv \sum_i \mathbf{p}_i(t) \, \delta[\mathbf{r}-\mathbf{r}_i(t)] \, ,\\
\label{Mkin}&&\hat{\kappa}(\mathbf{r},t) \equiv \sum_i \frac{|\mathbf{p}_i(t)|^2}{2\,m}\,\delta[\mathbf{r}-\mathbf{r}_i(t)] \,,
\end{eqnarray}
respectively, where $\delta(\mathbf{r})$ is the $d$-dimensional Dirac-delta distribution, and $d$ the spatial dimensionality of the problem. {The derivation of closed coarse-grained equations for slow variables starts with the derivation of the operator representation of the Hamiltonian equations, which is based on} the method of Zaccarelli et al  \cite{Zaccarelli2002} (also used by Archer  \cite{Archer_2006} and T\'oth  \cite{Toth_2020,TOTH2022133336}). {Accordingly, we calculate the Fourier transform of the microscopic densities, differentiate the result w.r.t. time, use the Fourier representation of the canonical equations, then calculate the inverse transform. The resulting equations read (for the details of the derivation, see Appendix A):}
\begin{eqnarray}
\label{continuity} \partial_t \hat{\rho} + \nabla\cdot\hat{\mathbf{g}} &=& 0\,,\\
\label{moment} \partial_t \hat{\mathbf{g}} + \nabla\cdot\hat{\mathbb{K}}&=& -\hat{\rho}\,{\nabla\hat{\Phi}}\,,\\
\label{thermal} \partial_t \hat{\kappa} + \nabla\cdot\hat{\mathbf{w}} &=& - \hat{\mathbf{g}} \cdot {\nabla\hat{\Phi}} \,,
\end{eqnarray}
where $\hat{\mathbb{K}}(\mathbf{r},t) = \sum_i \frac{\mathbf{p}_i(t)\,\otimes\, \mathbf{p}_i(t)}{m}\,\delta[\mathbf{r}-\mathbf{r}_i(t)]$ is the microscopic kinetic stress, {$\hat{\mathbf{w}}(\mathbf{r},t)=\sum_i \frac{|\mathbf{p}_i(t)|^2}{2\,m}\,\frac{\mathbf{p}_i(t)}{m}\,\delta[\mathbf{r}-\mathbf{r}_i(t)]$} the microscopic kinetic energy flux, {while
\begin{equation}
\label{Interaction}
\hat{\Phi}(\mathbf{r}) = \frac{1}{m}\,\sum_{n=1}^\infty\,\frac{1}{(n-1)!}\,\sum_{i_2,i_3,\dots,i_n} u_n(\mathbf{r},\mathbf{r}_{i_2},\mathbf{r}_{i_3},\dots,\mathbf{r}_{i_n})
\end{equation}
is the contribution of the particle interactions and the external field. 

To obtain general continuum equations from Eqns. (\ref{continuity})-(\ref{thermal}), one can utilise the method of ensemble averaging on Eqns. (\ref{continuity})-(\ref{thermal}) as follows.} Since the microscopic densities {defined by Eqns. (\ref{Mden})-(\ref{Mkin})} depend on the phase-space trajectory, {ensemble averaging} is based on calculating expectations over a random distribution of the initial conditions \cite{PhysRevLett.112.100602}. {Accordingly,} let
\begin{equation*}
\tilde{\mathbf{\Gamma}}(\mathbf{\Theta},t) \equiv (\mathbf{r}_1(\mathbf{\Theta},t),\mathbf{p}_1(\mathbf{\Theta},t),\mathbf{r}_2(\mathbf{\Theta},t),\mathbf{p}_2(\mathbf{\Theta},t),\dots,\mathbf{r}_N(\mathbf{\Theta},t),\mathbf{p}_N(\mathbf{\Theta},t))
\end{equation*}
denote the general solution of the Hamiltonian equations, where the positions and the momenta depend on time ($t$) and the initial conditions ($\mathbf{\Theta} \equiv (\mathbf{r}_1^0,\mathbf{r}_2^0,\dots,\mathbf{r}_N^0,\mathbf{p}_1^0,\mathbf{p}_2^0,\dots,\mathbf{p}_N^0)$ independently. {The microscopic densities of interest can be written in the generic form $\hat{A}(\mathbf{r},t,\mathbf{\Theta}) = \sum_i A[\mathbf{p}_i(\mathbf{\Theta},t)]\delta[\mathbf{r}-\mathbf{r}_i(\mathbf{\Theta},t)]$, where $A(\mathbf{p}_i)=m,\mathbf{p}_i,|\mathbf{p}_i|^2/(2\,m),$} etc. is the physical quantity carried by particle $i$. The ``coarse-grained'' counterpart of $\hat{A}(\mathbf{r},t,\mathbf{\Theta})$ is defined as {an expectation over a random distribution of the initial conditions, i.e.}: 
\begin{equation}
\label{CGIC} A(\mathbf{r},t) = \langle \hat{A}(\mathbf{r},t,\mathbf{\Theta}) \rangle_{0} \equiv \int_\Omega d\mathbf{\Theta}\, \hat{A}(\mathbf{r},t,\mathbf{\Theta})\,f_0(\mathbf{\Theta})\, ,
\end{equation}
where $\Omega$ denotes the phase-space, and $f_0(\Theta)$ is the p.d.f. of the initial conditions. {The meaning of Eq. (\ref{CGIC}) is as follows. Consider a trajectory starting from the phase-space point $\mathbf{\Gamma}=\mathbf{\Theta}$ at $t=0$, and measure the value of the physical quantity $A$ at time instant $t$. The contribution of $\hat{A}(\mathbf{r},t,\mathbf{\Theta})$ to the ensemble average is proportional to the probability that the system was at $\mathbf{\Gamma}=\mathbf{\Theta}$ at $t=0$, which is given by the p.d.f. $f_0(\mathbf{\Theta})$.} Since the integral w.r.t. $\Theta$ {in Eq. (\ref{CGIC})} does not affect $\mathbf{r}$ and $t$, applying Eq. (\ref{CGIC}) on Eqns. (\ref{continuity})-(\ref{thermal}) results in:
\begin{eqnarray}
\label{CG01}&&\partial_t \langle \hat{\rho} \rangle_0 + \nabla \cdot \langle \hat{\mathbf{g}} \rangle_0 = 0 \,,\\
\label{CG02}&&\partial_t \langle \hat{\mathbf{g}} \rangle_0 + \nabla \cdot \langle \hat{\mathbb{K}}  \rangle_0 = -\langle \hat{\rho}\,{\nabla\hat{\Phi}} \rangle_0 \,,\\
\label{CG03}&&\partial_t \langle\hat{\kappa}\rangle_0 + \nabla \cdot \langle\hat{\mathbf{w}}\rangle_0 = -\langle\hat{\mathbf{g}}\cdot {\nabla\hat{\Phi}}\rangle_0 \,.
\end{eqnarray}
{To express the coarse-grained quantities in terms of slow fields, we will approximate the solution of the Liouville equation. Since Eq. (\ref{CGIC}) contains the general solution of the Hamiltonian equations rather than that of the Liouville equation, first we prove that $A(\mathbf{r},t)$ defined by Eq. (\ref{CGIC}) is equivalent to the conventional ensemble average of the phase-space operator of $A$ over the solution of the Liouville equation. Introducing the phase-space coordinate $\mathbf{\Gamma}=(\mathbf{r}_1,\mathbf{p}_1,\dots,\mathbf{r}_2,\mathbf{p}_2,\dots,\mathbf{r}_N,\dots,\mathbf{p}_N)$, the conventional ensemble average of the physical quantity $A$ is defined as:}
\begin{equation}
\label{CGLE}
\langle \hat{A}(\mathbf{\Gamma},\mathbf{r}) \rangle_t \equiv \int_\Omega d\mathbf{\Gamma}\,\hat{A}(\mathbf{\Gamma},\mathbf{r})\,f(\mathbf{\Gamma}  ,t) \,,
\end{equation} 
where $\hat{A}(\mathbf{\Gamma},\mathbf{r})=\sum_i A(\mathbf{p}_i)\,\delta(\mathbf{r}-\mathbf{r}_i)$ is {the usual (time-independent) phase-space operator of $A$}, and $f(\mathbf{\Gamma},t)$ the solution of the Liouville equation. Following Khinchin \cite{khinchin1949mathematical}, the integral in Eq. (\ref{CGLE}) can be re-written {by considering the general solution of the Hamiltonian system as a variable transformation, i.e.: $\tilde{\mathbf{\Gamma}}: \Omega \times [0,+\infty) \to \Omega$ defined by $\mathbf{\Gamma}=\tilde{\mathbf{\Gamma}}(\mathbf{\Theta},t)$ assigns a point of the phase space to another one for any value of $t$.} Using the variable transformation $\mathbf{\Gamma}=\tilde{\mathbf{\Gamma}}(\mathbf{\Theta},t)$ in Eq. (\ref{CGLE}) yields: 
\begin{equation}
\label{CGinterim}
\langle \hat{A}(\mathbf{\Gamma},\mathbf{r}) \rangle_t = \int_{\Omega'} d\mathbf{\Theta}\, |\mathbb{J}| \,\hat{A}(\tilde{\mathbf{\Gamma}}(\mathbf{\Theta},t),\mathbf{r})\,f(\tilde{\mathbf{\Gamma}}(\mathbf{\Theta},t),t) \,.
\end{equation}
{$\hat{A}(\tilde{\mathbf{\Gamma}}(\mathbf{\Theta},t),\mathbf{r})$ means that we need to replace $\mathbf{r}_i$ and $\mathbf{p}_i$ by  $\mathbf{r}_i(\mathbf{\Theta},t)$ and $\mathbf{p}_i(\mathbf{\Theta},t)$, respectively, and therefore $\hat{A}(\tilde{\mathbf{\Gamma}}(\mathbf{\Theta},t),\mathbf{r}) = \hat{A}(\mathbf{r},t,\mathbf{\Theta})$ from Eq. (\ref{CGIC}). Since the phase-space flow is incompressible and $\tilde{\mathbf{\Gamma}}$ is a bijection (trajectories do not merge or split), $|\mathbb{J}|=1$ \cite{khinchin1949mathematical}) and $\Omega'=\Omega$. Furthermore, $f(\mathbf{\Gamma},t)$ is constant along trajectories, and therefore $f(\tilde{\mathbf{\Gamma}}(\mathbf{\Theta},t),t)=f(\tilde{\mathbf{\Gamma}}(\mathbf{\Theta},0),0)=f(\mathbf{\Theta},0)=f_0(\mathbf{\Theta})$}, where we used that $\tilde{\mathbf{\Gamma}}(\mathbf{\Theta},0)=\mathbf{\Theta}$ (by definition). Utilising these results in Eq. (\ref{CGinterim}) yields the following equivalence of ensemble averages:
\begin{equation}
\label{identity}
\langle \hat{A}(\mathbf{\Gamma},\mathbf{r}) \rangle_t = \langle \hat{A}(\tilde{\mathbf{\Gamma}}(\mathbf{\Theta},t),\mathbf{r}] \rangle_0 \,.
\end{equation}

{Eq. (\ref{identity}) is a central result here which allows us to replace the ensemble averages over the initial conditions by the conventional ones in Eqns. (\ref{CG01})-(\ref{CG03}), which results in the following general system of coarse-grained equations:}
\begin{eqnarray}
\label{CG1}\partial_t \langle \hat{\rho} \rangle_t+ \nabla \cdot \langle \hat{\mathbf{g}} \rangle_t &=& 0 \,,\\
\label{CG2}\partial_t \langle \hat{\mathbf{g}} \rangle_t+ \nabla \cdot \langle \hat{\mathbb{K}}  \rangle_t &=& -\langle \hat{\rho}\,{\nabla\hat{\Phi}} \rangle_t\,,\\
\label{CG3}\partial_t \langle\hat{\kappa}\rangle_t + \nabla \cdot \langle\hat{\mathbf{w}}\rangle_t &=& -\langle\hat{\mathbf{g}}\cdot {\nabla\hat{\Phi}}\rangle_t \,,
\end{eqnarray}
{where the microscopic densities are the following conventional phase-space operators}: $\hat{\rho}(\mathbf{\Gamma},\mathbf{r}) = \sum_i m\,\delta(\mathbf{r}-\mathbf{r}_i)$, $\hat{\mathbf{g}}(\mathbf{\Gamma},\mathbf{r}) = \sum_i \mathbf{p}_i\,\delta(\mathbf{r}-\mathbf{r}_i)$, $\hat{\mathbb{K}}(\mathbf{\Gamma},\mathbf{r}) = \sum_i \frac{\mathbf{p}_i \otimes \mathbf{p}_i}{m}\,\delta(\mathbf{r}-\mathbf{r}_i)$, $\hat{\kappa}(\mathbf{\Gamma},\mathbf{r})=\sum_i \frac{|\mathbf{p}_i|^2}{2\,m}\,\delta(\mathbf{r}-\mathbf{r}_i)$, and $\hat{\mathbf{w}}(\mathbf{\Gamma},\mathbf{r})=\sum_i \frac{|\mathbf{p}_i|^2\mathbf{p}_i}{2\,m^2}\delta(\mathbf{r}-\mathbf{r}_i)$. The practical importance of Eq. (\ref{identity}) is bifold: (i) It is a ``shortcut'' to the coarse-grained equations, since it makes the utilisation of Green's integral theorem unnecessary when transferring the phase-space derivatives w.r.t. the particle position $\mathbf{r}_i$ to real-space ones w.r.t. $\mathbf{r}$ \cite{doi:10.1063/1.1747782}, and, (ii) it allows us to calculate the coarse-grained quantities for non-equilibrium solutions of the Liouville equation. {Since exact solutions are unknown, approximating ones parametrised by slow fields must be constructed following physical principles, and therefore the theory presented in the rest of the paper can be interpreted as a semi-phenomenological approach.}

\section{Dissipative adiabatic fluids}

\subsection{Local Thermodynamic Equilibrium}

{To obtain closed hydrodynamic equations from Eqns. (\ref{CG1})-(\ref{CG3}) for slow fields describing the macroscopic physical state of the system, the solution of the Liouville equation is approximated as follows. {Since the particles are identical, the reduced many-body p.d.f. defined by
\begin{equation}
\label{reducedpdf}
f^{(n)}(\mathbf{r}_{i_1},\mathbf{p}_{i_1},\mathbf{r}_{i_2},\mathbf{p}_{i_2},\dots,\mathbf{r}_{i_n},\mathbf{p}_{i_n},t) \equiv \int \prod_{k=n+1}^{N} d\mathbf{r}_{i_k}\,d\mathbf{p}_{i_k}\,f(\mathbf{\Gamma},t)
\end{equation}
(where $i_k \in (1,2,\dots,N)$, $k=1,2,\dots N$ and $i_k \neq i_l$ for $k \neq l$, $l=1,2,\dots,N$)} is invariant for the permutation of the particle indices $(i_1,i_2,\dots,i_n)$ for $n=1,2,\dots N$. Consequently, the $n$-body p.d.f's are identical. Extending Kreuzer's approach \cite{Kreuzer} to the non-isothermal case, the unique single-particle p.d.f. is approximated as:
\begin{equation}
\label{f1thermal}
f^{(1)}(\mathbf{r}_i,\mathbf{p}_i,t) \approx  f^{(1)}_0(\mathbf{r}_i,\mathbf{p}_i,t)- f^{(1)}_v(\mathbf{r}_i,\mathbf{p}_i,t) - f^{(1)}_T(\mathbf{r}_i,\mathbf{p}_i,t) \,,
\end{equation} 
where
\begin{equation}
\label{LTE}
f^{(1)}_0(\mathbf{r},\mathbf{p},t) = \frac{\rho(\mathbf{r},t)}{M}\,\frac{1}{(2\pi\,m\,k_B \,T(\mathbf{r},t))^{d/2}}\,\exp\left({-\frac{|\mathbf{p}-m\,\mathbf{v}(\mathbf{r},t)|^2}{2\,m\,k_B T(\mathbf{r},t)}}\right)
\end{equation} 
refers to Local Thermodynamic Equilibrium (LTE) characterised by the macroscopic slow fields $\rho(\mathbf{r},t)$ (mass density), $\mathbf{v}(\mathbf{r},t)$ (velocity), and $T(\mathbf{r},t)$ (temperature) \cite{hansen90a}, while the leading-order dissipative corrections read:
\begin{eqnarray*}
f^{(1)}_v(\mathbf{r},\mathbf{p},t) &=& A(\lambda)\left( \mathbf{n}\cdot\mathbb{V}\cdot\mathbf{n} - \frac{1}{3}\,\nabla \cdot\mathbf{v} \right) \,,\\
f^{(1)}_T(\mathbf{r},\mathbf{p},t) &=& B(\lambda)\,(\mathbf{n}\cdot\mathbf\nabla T) \,,
\end{eqnarray*}
where $\lambda$ is the magnitude and $\mathbf{n}$ the direction of the relative particle momentum $\mathbf{x} \equiv \mathbf{p}-m\,\mathbf{v}(\mathbf{r},t)$, while $\mathbb{V} = \frac{1}{2}\left[ (\nabla \otimes \mathbf{v}) + (\nabla \otimes \mathbf{v})^T \right]$. (Higher-order corrections containing the spatio-temporal derivatives of the slow fields can be added to Eq. (\ref{f1thermal}) to include non-adiabatic effects \cite{10.1063/1.4807586,RevModPhys.94.015007}, but these are not considered here.) Using Eq. (\ref{f1thermal}) in Eqns. (\ref{CG1})-(\ref{CG3}) results in (for the details of the calculation, see Appendix B):
\begin{eqnarray}
\label{Appdiss1}\partial_t \rho + \nabla \cdot(\rho\,\mathbf{v}) &=& 0 \,,\\
\label{Appdiss2}\partial_t(\rho\,\mathbf{v}) + \nabla\cdot(\rho\,\mathbf{v}\otimes\mathbf{v}) + \nabla p_0  - \nabla\cdot(\eta\,\mathbb{D})&=& - \langle\hat{\rho}\,\nabla\hat{\Phi}\rangle_t\,,\\
\label{Appdiss3}\partial_t \kappa + \nabla\cdot[(\kappa+p_0)\mathbf{v}]- \nabla\cdot\,\mathbf{J} &=& -\langle \hat{\mathbf{g}}\cdot\nabla\hat{\Phi} \rangle_t \,,
\end{eqnarray} 
where $p_0 = (k_B T/m)\,\rho$ is the ideal gas pressure, $\eta$ the dynamic viscosity, $\mathbb{D}(\mathbf{v})=(\nabla \otimes \mathbf{v})+(\nabla \otimes \mathbf{v})^T - \frac{2}{3}\,(\nabla\cdot\mathbf{v})\,\mathbb{I}$ the viscous stress (where $\mathbb{I}$ is the $d \times d$ identity matrix), $\kappa=q+\lambda$ the total kinetic energy density with $q=\frac{d}{2}\,p_0$ being the thermal energy density, and $\lambda=(1/2)\,\rho\,|\mathbf{v}|^2$ the macroscopic kinetic energy density, while the dissipative energy flux reads: $\mathbf{J}=\eta\,\mathbb{D}\cdot\mathbf{v}+\chi\,\nabla T$, where $\chi$ stands for the thermal conductivity. For $\eta=\chi=\hat{\Phi}=0$, Eqns. (\ref{Appdiss1})-(\ref{Appdiss3}) reduce to the phenomenological hydrodynamic equations for non-isothermal ideal gases that can be obtained in the framework of the Kinetic Theory. Therefore, one can conclude that the principle of Local Thermodynamic Equilibrium for a non-interacting system defines the non-equilibrium ideal gas.}

\subsection{Independence of positions and momenta}

{To model physical phenomena beyond the ideal gas, the interaction terms in Eqns. (\ref{Appdiss2}) and (\ref{Appdiss3}) must be expressed in terms of the slow fields. Using Eqns. (\ref{Mden}), (\ref{Interaction}) and (\ref{CGLE}) in Eq. (\ref{Appdiss2}) yields:
\begin{equation}
\label{interNS}
\langle\hat{\rho}\, \nabla \Phi\rangle_t = N\,\sum_{n=1}^\infty {N-1 \choose n-1}\int \prod_{k=2}^n d\mathbf{r}_k\, \nabla u_n(\mathbf{r},\mathbf{r}_2,\dots,\mathbf{r}_n)\,\varrho^{(n)}(\mathbf{r},\mathbf{r}_2,\dots,\mathbf{r}_n,t) \,,
\end{equation}
where $\varrho^{(n)}(\mathbf{r}_1,\mathbf{r}_2,\dots,\mathbf{r}_n,t) = \,\int \prod_{k=1}^n d\mathbf{p}_{k}\,f^{(n)}(\mathbf{r}_1,\mathbf{p}_1,\mathbf{r}_2,\mathbf{p}_2,\dots,\mathbf{r}_n,\mathbf{p}_n,t)$ is the reduced $n$-particle density. Furthermore, using Eqns. (\ref{Mmom}), (\ref{Interaction}) and (\ref{CGLE}) in Eq. (\ref{Appdiss3}) results in:
\begin{equation}
\label{interkin}
\begin{split}
\langle\hat{\mathbf{g}}\cdot\nabla\hat{\Phi}\rangle_t  =  & \frac{N}{m}\, \sum_{n=1}^\infty {N-1 \choose n-1} \int \prod_{k=2}^n d\mathbf{r}_k d\mathbf{p}_k\, \nabla u_n(\mathbf{r},\mathbf{r}_2,\dots,\mathbf{r}_n)\\
& \times \int d\mathbf{p}\,\mathbf{p}\,f^{(n)}(\mathbf{r},\mathbf{p},\mathbf{r}_2,\mathbf{p}_2,\dots,\mathbf{r}_n,\mathbf{p}_n,t) \,,
\end{split}
\end{equation}
where $f^{(n)}(.)$ is defined by Eq. (\ref{reducedpdf}). By comparing Eqns. (\ref{interNS}) and (\ref{interkin}), it is easy to see that requiring 
\begin{equation}
\label{Fcond}
\int \prod_{k=1}^n d\mathbf{p}_k\, \mathbf{p}_i\, f^{(n)}(\mathbf{r}_1,\mathbf{p}_1,\mathbf{r}_2,\mathbf{p}_2,\dots,\mathbf{r}_n,\mathbf{p}_n,t) = m\,\mathbf{v}(\mathbf{r}_i,t)\,\varrho^{(n)}(\mathbf{r}_1,\mathbf{r}_2,\dots,\mathbf{r}_n,t)
\end{equation} 
for $i=1,2,\dots,n$ and $n=2,3,\dots,N$ results in
\begin{equation}
\label{Tcond}
\langle\hat{\mathbf{g}}\cdot\nabla\hat{\Phi}\rangle_t = \mathbf{v}(\mathbf{r},t)\,\langle\hat{\rho}\cdot\nabla\hat{\Phi}\rangle_t \,,
\end{equation} 
which allows us to proceed with the derivation of the coarse-grained equations as follows. If Eq. (\ref{Tcond}) holds, the interaction term can be eliminated from Eq. (\ref{Appdiss3}) by using Eq. (\ref{Appdiss2}). Furthermore, using $\kappa = \frac{1}{2}\rho\,|\mathbf{v}|^2 + \frac{d}{2}\,\frac{k_B T}{m}\,\rho$ in the resulting equation yields the following closed equation for the temperature (for the details of the derivation, see Appendix C):
\begin{equation}
\label{AppTdiss}
\frac{d}{2}\left[ \partial_t \ln\left(\frac{T}{\tau}\right) + \mathbf{v}\cdot\nabla \ln\left(\frac{T}{\tau} \right) \right] = - \nabla\cdot\mathbf{v} +\frac{D}{p_0} \, ,
\end{equation}
where $\tau >0$ is an arbitrary reference temperature, and $D=\nabla\cdot(\chi\,\nabla T)+\eta\,\mathbb{D}:(\nabla\otimes\mathbf{v})$ accounts for heat diffusion and generation due to internal friction. We emphasize here that Eq. (\ref{AppTdiss}) is a direct consequence of Eq. (\ref{Fcond}), and does not depend on the details of the interactions. Whether Eq. (\ref{Fcond}) is a harsh condition or not can be easily found out using conditional probability as follows. The expectation of the momentum of particle $i$ given that particle 1 is at position $\mathbf{r}_1$, particle 2 is at position $\mathbf{r}_2$, etc. reads as (by definition):
\begin{equation}
\label{condexp}
\mathbb{E}(\mathbf{p}_i\,|\,\mathbf{r}_1,\mathbf{r}_2,\dots,\mathbf{r}_n) = \int \prod_{k=1}^n d\mathbf{p}_k\,\mathbf{p}_i\,\frac{f^{(n)}(\mathbf{r}_1,\mathbf{p}_1,\mathbf{r}_2,\mathbf{p}_2,\dots,\mathbf{r}_n,\mathbf{p}_n,t)}{\varrho^{(n)}(\mathbf{r}_1,\mathbf{r}_2,\dots,\mathbf{r}_n,t)} \,,
\end{equation}
where the expectation also refers to time instant $t$. If Eq. (\ref{Fcond}) holds, Eq. (\ref{condexp}) reduces to
\begin{equation}
\label{selfconsistent}
\mathbb{E}(\mathbf{p}_i|\,\mathbf{r}_1,\mathbf{r}_2,\dots,\mathbf{r}_n) = m\,\mathbf{v}(\mathbf{r}_i,t)
\end{equation}
for $i=1,2,\dots,n$ and $n=1,2,\dots,N$, thus expressing that the expectation of the momentum of particle $i$ only depends on the position of the particle, and is given by the slow velocity field. If Eq. (\ref{selfconsistent}) applies, the phenomenological theory is recovered.}

\subsection{Thermodynamic consistency}

{Since Eqns. (\ref{Appdiss1}) and (\ref{AppTdiss}) are already closed for the macroscopic fields, the remaining step is to express the interaction term in Eq. (\ref{Appdiss2}) in terms of the slow fields. Accordingly, first we re-write the gradient of the ideal gas pressure as
$\nabla p_0 = \rho\,\nabla(\delta_\rho F_{id})-(\delta_T F_{id})\,\nabla T$, where $F_{id}[\rho,T] = \int d\mathbf{r}\,\frac{k_B T}{m}\rho\left[\log\left(\frac{\rho}{\delta}\right)-1 \right]$ is {associated with} the free energy of the ideal gas (where $\delta>0$ is an arbitrary reference density), and $\delta_\rho F_{id}$ and $\delta_T F_{id}$ stand for the first functional derivative of $F_{id}[\rho,T]$ w.r.t. $\rho$ and $T$, respectively. Then,  following Kikkinides and Monson \cite{doi:10.1063/1.4913636}, we assume that there exists $F_{exc}[\rho,T]=\int d\mathbf{r}\,u$ for which
\begin{equation}
\label{Fexc}
\langle \hat{\rho}\,\nabla \hat{\Phi} \rangle_t = \rho\,\nabla(\delta_\rho F_{exc})-(\delta_T F_{exc})\, \nabla T
\end{equation}
holds, which is merely the non-isothermal extension of the adiabatic theory. If Eq. (\ref{Fexc}) holds, Eq. (\ref{Appdiss2}) can be re-written in variational form as:
\begin{equation}
\label{varNS}
\partial_t(\rho\,\mathbf{v}) + \nabla\cdot(\rho\,\mathbf{v}\otimes\mathbf{v}) = -\rho\,\nabla(\delta_\rho F) + (\delta_T F)\,\nabla T + \nabla\cdot(\eta\,\mathbb{D}) \,,
\end{equation}
where the Helmholtz free energy of the system reads:
\begin{equation}
\label{freeenergy}
F[\rho,T] = F_{id}[\rho,T] + F_{exc}[\rho,T] + F_{g}[\rho,T] \,.
\end{equation}
Here $F_g[\rho,T]$ is a gauge-invariant term satisfying the equation 
\begin{equation}
\label{gcond}
(\delta_T F_{g})\,\nabla T = \rho\,\nabla (\delta_\rho F_{g}) \,,
\end{equation}
and therefore $F_g[\rho,T]$ provides no contribution to the dynamical equations.}

{To show that Eq. (\ref{Fexc}) is thermodynamically consistent, one needs to prove that entropy reaches its maximum at equilibrium (note that the system is isolated). According to Eqns. (\ref{Appdiss1}), (\ref{AppTdiss}) and (\ref{varNS}), equilibrium is characterised by $\mathbf{v}(\mathbf{r},t)=\mathbf{0}$ (no flow), $\delta_\rho F=\mu_0$ (constant chemical potential), and $T=T_0$ (constant temperature), at which entropy must take its maximum (at constant mass and energy). The non-equilibrium entropy and internal energy of the system can be written as $S = \int d\mathbf{r}\,s$ and $E=\int d\mathbf{r}\,e$, respectively, where $s =-\delta_T F$ and $e = f + T\,s$ (Legendre transform) are the non-equilibrium entropy density and internal energy density, respectively. Since $\mathbf{v}(\mathbf{r},t)=\mathbf{0}$ at equilibrium, the total energy density defined by $\epsilon \equiv \frac{1}{2}\,\rho\,|\mathbf{v}^2|+e$ and the internal energy density coincide there. Accordingly, using the Lagrange multiplier method, the following Euler-Lagrange equations should apply at equilibrium:
\begin{eqnarray}
\label{thermoEL1} \delta_\rho S - \xi\,(\delta_\rho E) - \gamma\,(\delta_\rho M) &=& 0 \,, \\
\label{thermoEL2} \delta_T S - \xi\,(\delta_T E) - \gamma\,(\delta_T M) &=& 0 \,,
\end{eqnarray}
where $M = \int d\mathbf{r}\,\rho$ is the mass of the system, while $\xi$ and $\gamma$ are (constant) Lagrange multipliers. Setting $T \equiv T_0$ (constant) and $\xi \equiv 1/T_0$, and using $\delta_\rho M = 1$ and $(\delta_\rho E)_{T=T_0} = (\delta_\rho F)_{T=T_0}+T_0\,(\delta_\rho S)_{T=T_0}$ in Eq. (\ref{thermoEL1}), then multiplying by $T_0$ yields: $(\delta_\rho F)_{T=T_0} = \gamma\,T_0$ (constant). Furthermore, Eq. (\ref{thermoEL2}) is also satisfied for $T=T_0$ and $\xi=1/T_0$, since $\delta_T M =0$, and $(\delta_T E)_{T=T_0} = (\delta_T F)_{T=T_0} + s_{T=T_0} + T_0\,(\delta_T S)_{T=T_0} = T_0\,(\delta_T S)_{T=T_0}$. Consequently, Eqns. (\ref{thermoEL1}) and (\ref{thermoEL2}) are identical to $T=T_0$ and $\delta_\rho F=\mu_0$ at $\mathbf{v}(\mathbf{r},t)=\mathbf{0}$.}

{Although the conditional stationary points of the entropy coincide with the stationary solutions of the dynamical equations, we still need to show that equilibrium occurs at the maximum of the entropy. In general, if there exists $F_g[\rho,T]$ for which the entropy production rate is non-negative, the dynamics finds the maximum of the entropy. For the ideal gas, we assume $F_g[\rho,T] = \int d\mathbf{r}\,f_g^{(id)}(\rho,T)$, where the integrand is a local function of $\rho$ and $T$. Using Eq. (\ref{gcond}) then results in the general form $f^{(id)}_g(\rho,T)=\rho\,h_{id}(T)+C$, where $C$ is a constant. Finally, using $s = -\frac{\partial f_{id}}{\partial T} - \rho \frac{dh_{id}}{dT}$ and $e=q=\frac{d}{2}\,\frac{k_B T}{m}\,\rho$ (the internal energy density coincides with the thermal part of the kinetic energy density) in the Legendre transform $e=f+T\,s$ yields
\begin{equation}
\label{hid}
h_{id}(T) = \frac{d}{2}\,\frac{k_B T}{m}\,\left[ 1-\ln\left(\frac{T}{\tau} \right)  \right] \,.
\end{equation}
Using Eq. (\ref{hid}), the entropy production rate for a general interacting system reads:
\begin{equation}
\label{AppSfull}
\dot{S} = \frac{k_B}{m}\,\frac{d}{dt} \int d\mathbf{r} \,\rho\, \left[1- \ln\left(\frac{\rho}{\delta} \right)+\frac{d}{2}\,\ln\left( \frac{T}{\tau}\right) \right] - \frac{d}{dt}\int d\mathbf{r}\,(\delta_T F_{exc}) \,.
\end{equation}
It can be shown that the first integral is related to dissipation as (for details, see Appendix C):
\begin{equation}
\label{Sprod}
\dot{S} = \int d\mathbf{r}\,\left( \kappa\,\left| \frac{\nabla T}{T} \right|^2 + \frac{\eta}{2\,T} (\Sigma\mathbb{D})^2\right) - \frac{d}{dt}\int d\mathbf{r}\,(\delta_T F_{exc} ) \,,
\end{equation}  
where the first term is non-negative and agrees with the well-known result of Landau and Lifshitz \cite{Landau1987Fluid}. The second law of thermodynamics states that non-equilibrium processes in isolated systems are entropy monotonous. Although one might argue that $F_g[\rho,T]$ should be extended with a term that compensates the second integral in Eq. (\ref{Sprod}), Eq. (\ref{gcond}) might be too stringent, since, contrary to the free energy of the ideal gas, $F_{exc}[\rho,T]$ can be arbitrary. Consequently, the existence of $F_g[\rho,T]$ for which the second integral in Eq. (\ref{Sprod}) is constant cannot be guaranteed in general. For this reason, we limit ourselves to scenarios where the temperature dependence of the excess free energy can be neglected. Accordingly, the general form of the free energy for which the entropy production rate in a dissipative and non-isothermal isolated system is surely non-negative reads: 
\begin{equation}
\label{genF}
F[\rho,T ]= \int d\mathbf{r} \, \left\{ p_0 \left[ \ln\left(\frac{\rho}{\delta}\right)-1 \right] + u + \rho\,h_{id}(T) \right\} \,,
\end{equation}
where $h_{id}(T)$ is defined by Eq. (\ref{hid}), and $u$ is a non-local function of $\rho$.} {The physical meaning of $h_{id}(T)$ can be found out by re-formulating the integrand of $F[\rho,T]$ by using Eq. (\ref{hid}), thus yielding:
\begin{equation}
\label{deBroglie}
F[\rho,T ]= \int d\mathbf{r} \, \left\{ p_0\left[ \ln\left(\frac{\rho}{m}\,\Lambda^d\right)-1 \right] + e \right\} \,,
\end{equation}
where the first term of the sum is the free energy density of the ideal gas, while $e = u+\frac{d}{2}\,p_0$ is the internal energy density. Furthermore, $\Lambda = \left(\frac{m}{\delta}\right)^{1/d}\,\left(\frac{\tau}{T}\right)^{1/2}$ is the thermal de Broglie wavelength (for suitable choices of $\delta$ and $\tau$). Eq. (\ref{genF}) then expresses that even though the de Broglie wavelength gives no contribution to the dynamics, its temperature dependence is necessary for the non-negativity of the entropy production rate.} 

\section{Non-local energy transport}

Since $u$ is non-local in Eq. (\ref{genF}), it is worth to address the possible modes of energy transport in dissipative adiabatic fluids. Our starting point is Eq. (\ref{Appdiss3}) when Eqns. (\ref{Tcond}), (\ref{Fexc}) and (\ref{genF}) also apply:
\begin{equation}
\label{kin1}
\partial_t \kappa + \nabla\cdot[(\kappa+p_0)\,\mathbf{v}] - \nabla \cdot \mathbf{J} = -\rho\,\mathbf{v} \cdot \nabla(\delta_\rho F_{exc}) \enskip .
\end{equation}
The total energy density of the system is $\epsilon = \lambda + e = \lambda + (q + u) = \kappa + u$, where we used $e=f+T\,s$. Furthermore, the pressure reads as $p = \rho\,(\delta_\rho F)-f=p_0+\rho\,(\delta_\rho F_{exc})-u$. Adding $\partial_t u+\nabla\cdot[\rho\,\mathbf{v}\,(\delta_\rho F_{exc})]$ to both sides of Eq. (\ref{kin1}), then using the continuity equation results in:
\begin{equation}
\label{energybalance}
\partial_t \epsilon + \nabla\cdot[(\epsilon+p)\,\mathbf{v}] - \nabla\cdot\mathbf{J} = \sigma_{exc}  \,,
\end{equation}
where $\sigma_{exc} = \partial_t u - (\partial_t \rho)(\delta_\rho F_{exc})$ is associated with the particle interactions. If the integrand of the excess free energy can be written as a function of the density and its spatial derivatives, $\sigma_{exc}$ can be expressed as:
\begin{equation}
\label{nonlocalexpand}
\sigma_{exc} =  \sum_{i=1}^\infty \left( \frac{\partial u}{\partial \nabla^i \rho} \cdot \nabla^i(\partial_t \rho) - (-1)^i\,(\partial_t \rho)\,\nabla^i\,\frac{\partial u}{\partial \nabla^i \rho} \right) \,.
\end{equation} 
Using Eq. (\ref{nonlocalexpand}), it is simple to show that $\int d\mathbf{r}\,\sigma_{exc} =0$, and therefore the total energy ($E_{tot} \equiv \int d\mathbf{r}\,\epsilon$) is conserved despite that fact that $\sigma_{exc}$ is a non-local source term. Regarding the role of $\sigma_{exc}$, it is trivial to see that $\sigma_{exc}=0$ for local theories (i.e., when the excess free energy only depends on the density but not on its derivatives). In this case, Eq. (\ref{energybalance}) reduces to $\partial_t \epsilon + \nabla\cdot[(\epsilon+p)\,\mathbf{v}]=0$, which is identical to the energy balance equation derived in the framework of the phenomenological theory \cite{Landau1987Fluid,hansen90a} or the Kinetic Theory \cite{chapman1990mathematical}. The dominance of locality is also expected for incompressible liquids and dilute gases for the following reasons. In the former, the density is constant, and therefore $\partial_t \rho=0$, while in the latter, the dimensionless equations approximate the kinetic limit: When the length-scale $\Lambda \equiv (m/\bar{\rho})^{1/d}$ (where $\bar{\rho}$ is the average mass density of the system) is much larger than the characteristic length-scale of the interaction potential(s), a local theory is recovered \cite{TOTH2022133336}, and therefore the dimensionless counterpart of $\sigma_{exc}$ will tend to zero. $\sigma_{exc}$ is expected to be dominant in case of high density and compressibility, which is characteristic to near-critical fluids. {Equilibrium gas-liquid interfaces are extensively studied in the framework of the Classical Density Functional Theory, for instance. Accordingly, the excess free energy density of an inhomogeneous single-component system is written as \cite{doi:10.1080/00018737900101365,hansen90a,doi:10.1080/00018732.2020.1854965}:
\begin{equation}
\label{CDFT}
u = \rho(\mathbf{r})\left(\frac{V_{ext}(\mathbf{r})}{m} -k_B T_0\,\sum_{n=2}^\infty \frac{1}{n!}\int \prod_{i=1}^{n-1} d\mathbf{r}_i\,\rho(\mathbf{r}_i)\,C_n(\mathbf{r},\mathbf{r}_1,\mathbf{r}_2,\dots,\mathbf{r}_{n-1}) \right) \,,
\end{equation}
where $V_{ext}(\mathbf{r})$ is the external field, $T_0$ a (constant) reference temperature, and $C_n(\dots)$ is related to the $n$-point correlation function at $T_0$. The excess free energy defined by Eq. (\ref{CDFT}) is non-local and only depends on the density, therefore it belongs to a thermodynamically consistent family of theories defined here. For a suitable choice of the correlation functions, the theory provides a gas-liquid critical point. Near the critical point, $\sigma_{exc}$ becomes significant, and due to its non-local nature,} a faster-than-convection transport mode of energy is expected to emerge.

\section{Concluding remarks}

{We presented a statistical mechanical derivation of thermodynamically consistent fluid mechanical equations for simple non-isothermal dissipative fluids. The derivation started with the operator representation of the Hamiltonian particle dynamics. From the exact equations for the microscopic densities, closed coarse-grained equations were obtained for the density, temperature, and velocity fields in 3 major steps as follows.}

\noindent{(i) The principle of Local Thermodynamic Equilibrium (LTE) was used to obtain hydrodynamic equations for the non-isothermal ideal gas. First-order corrections around LTE were considered to include dissipative terms for adiabatic fluids.}

\noindent{(ii) We assumed that the conditional expectations of the momentum of a particle coincide, and only depend on the position of the particle in a way that the velocity is picked from the hydrodynamic velocity field.}

\noindent{(iii) We assumed that the coarse-grained interaction term in the equations can be written in variational form, thus providing a non-isothermal extension of the conventional Dynamical Density Functional Theory.}

{Applying the principle of LTE led to 3 coupled equations, out of which only the continuity equation was closed for the coarse-grained variables. Using the assumption of independent position and momenta led to a second closed coarse-grained equation for the temperature. Finally, the assumption of the variational formulation provided (formally) closed Navier-Stokes equations. The final equations of motion read:
\begin{eqnarray}
\partial_t \rho + \nabla\cdot(\rho\,\mathbf{v}) &=& 0 \,,\\
\partial_t(\rho\,\mathbf{v})+\nabla\cdot(\rho\,\mathbf{v}\otimes\mathbf{v}) &=& {-\nabla p_0 -\rho\,\nabla(\delta_\rho F_{exc})}+\nabla\cdot(\eta\,\mathbb{D}) \,,\quad\quad\\
\frac{d}{2}\left(\partial_t \ln\left( \frac{T}{\tau} \right) + \mathbf{v}\cdot\nabla \ln \left( \frac{T}{\tau}\right) \right) &=& -\nabla\cdot\mathbf{v} + \frac{D}{p_0} \,,  
\end{eqnarray}
{where $p_0=(k_B T/m)\,\rho$ is the ideal gas pressure and $F_{exc}[\rho]=\int d\mathbf{r}\,u$ is the excess free energy (including the external field), where, in case of a CDFT-based free energy, $u$ is defined by Eg. (\ref{CDFT}).} Furthermore, $\mathbb{D}$ is the viscous stress, $p_0$ the ideal gas pressure, and $D=\eta\,\mathbb{D}:(\nabla\otimes\mathbf{v})+\nabla\cdot(\chi\,\nabla T)$ is responsible for heat generation and diffusion. The free energy given by Eq. (\ref{genF}) defines a functional class for which the conventional thermodynamic formalism can be adopted for non-equilibrium situations without further corrections. The pressure, the chemical potential, the internal energy density, and even the entropy density are all well-defined in non-equilibrium, while the dynamics is entropy monotonic for an isolated system. Although Eq. (\ref{genF}) provides no ultimate resolution to the problem of the consistent non-equilibrium generalisation of thermodynamics, it offers a flexible framework for practical applications. As another major achievement of the paper, we have shown that the non-local nature of the particle interactions contribute to energy transport. The energy balance equation defined by Eq. (\ref{energybalance}) contains a non-local source term which is negligible in incompressible fluids and dilute gases, but may become dominant in near-critical fluids. In reality, a faster-than-convection mode of heat transport is present in these fluids, a phenomenon which serves as the basis of an industrial technology for efficient cooling of nuclear reactors. We expect that the phenomenon can be modelled and understood within the framework of the theory presented here.}

\color{black}

\section*{Acknowledgements}
The author would like to thank Isaac Newton Institute for Mathematical Sciences for support and hospitality during the programme ``New statistical physics in living matter: non equilibrium states under adaptive control'' when work on this paper was undertaken. This work was supported by: EPSRC Grant Number EP/R014604/1.

\section*{Appendix A. Derivation of the microscopic continuum equations}

The Fourier transform of the microscopic mass, momentum, and kinetic energy density read:
\begin{eqnarray}
\label{FourP}\hat{P}(\mathbf{k},t) &=& \frac{1}{(2\pi)^d}  \sum_i m \, e^{-\imath\,\mathbf{k}\cdot\mathbf{r}_i(t)} \,, \\
\label{FourG}\hat{\mathbf{G}}(\mathbf{k},t) &=& \frac{1}{(2\pi)^d} \sum_i \mathbf{p}_i(t) e^{-\imath\,\mathbf{k}\cdot\mathbf{r}_i(t)}  \,, \\
\label{FourK}\hat{N}(\mathbf{k},t) &=& \frac{1}{(2\pi)^d} \sum_i \frac{|\mathbf{p}_i(t)|^2}{2\,m} e^{-\imath\,\mathbf{k}\cdot\mathbf{r}_i(t)}  \,,
\end{eqnarray}
respectively, where the Fourier transform of $f(\mathbf{r})$ is defined as $$F(\mathbf{k}) \equiv \frac{1}{(2\,\pi)^d} \int d\mathbf{r}\,f(\mathbf{r})\,e^{-\imath\,\mathbf{k}\cdot\mathbf{r}}\,,$$ and $\mathbf{r} \in \mathbb{R}^d$ with $d=1,2,3$. For the sake of compactness of the formulae, we omit the argument $t$ in $\mathbf{r}_i(t)$ and $\mathbf{p}_i(t)$ henceforth. Differentiating $\hat{P}(\mathbf{k},t)$ w.r.t. $t$ yields:
\begin{equation}
\label{derP}
\begin{split}
\partial_t \hat{P}(\mathbf{k},t) & = \frac{1}{(2\pi)^d} \sum_i m\, ( -\imath\,\mathbf{k} \cdot\dot{\mathbf{r}}_i) \, e^{-\imath\,\mathbf{k}\cdot\mathbf{r}_i} \\ 
& = -\imath\,\mathbf{k} \cdot \frac{1}{(2\pi)^d}  \sum_i \mathbf{p}_i \, e^{-\imath\,\mathbf{k}\cdot\mathbf{r}_i} = -\imath\,\mathbf{k} \cdot \hat{\mathbf{G}}(\mathbf{k},t) \,,
\end{split}
\end{equation}
where we used the canonical equations. The inverse Fourier transform of Eq. (\ref{derP}) yields the microscopic continuity equation:
\begin{equation}
\label{continuityA}
\partial_t \hat{\rho} + \nabla\cdot\hat{\mathbf{g}} = 0\,.
\end{equation}
Similarly to the mass density, one can differentiate $\hat{\mathbf{G}}(\mathbf{k},t)$ w.r.t. $t$, which gives:
\begin{equation}
\label{derG} 
\partial_t \hat{\mathbf{G}}(\mathbf{k},t) = \frac{1}{(2\pi)^d} \sum_i \left[\mathbf{p}_i \left(-\imath\,\mathbf{k}\cdot\frac{\mathbf{p}_i}{m}\right) - \frac{\partial \hat{V}}{\partial \mathbf{r}_i} \right] e^{-\imath\,\mathbf{k}\,\mathbf{r}_i}
\end{equation}
where we used the canonical equations. The potential energy of the system reads:
\begin{equation}
\hat{V} = \sum_i u_1(\mathbf{r}_i) + \frac{1}{2!}\sum_{i,j} u_2(\mathbf{r}_i,\mathbf{r}_j)+\frac{1}{3!}\sum_{i,j,k} u_3(\mathbf{r}_i,\mathbf{r}_j,\mathbf{r}_k) + \dots \,,
\end{equation}
where $u_n(.)=0$ whenever any two of its arguments coincide. We assume that the many body potentials can be represented as
\begin{equation}
\label{Fourierrep}
u_n(\mathbf{r}_1,\mathbf{r}_2,\dots,\mathbf{r}_n) = \int d\mathbf{q}\,U_n^{(i)}(\mathbf{q},\{\mathbf{r}_k\})\,e^{\imath\,\mathbf{q}\cdot\mathbf{r}_i} \,,
\end{equation}
where 
\begin{equation}
U_n^{(i)}(\mathbf{q},\{\mathbf{r}_k\}) = \frac{1}{(2\pi)^d} \int d\mathbf{r}_i\,u_n(\mathbf{r}_1,\mathbf{r}_2,\dots,\mathbf{r}_n)\,e^{-\imath\,\mathbf{q}\cdot\mathbf{r}_i}
\end{equation}
is the Fourier transform of $u_n(.)$ w.r.t. its $i^\mathrm{th}$ argument (where $i=1,2,\dots,n$), and therefore $\{\mathbf{k}_r\}=(\mathbf{r}_1,\mathbf{r}_2,\dots,\mathbf{r}_{i-1},\mathbf{r}_{i+1},\dots,\mathbf{r}_n)$, i.e., $U_n^{(i)}(\mathbf{q},\{\mathbf{r}_k\})$ does not depend on $\mathbf{r}_i$. The derivative of $\hat{V}$ w.r.t. $\mathbf{r}_i$ reads:
\begin{equation}
\label{potentialexp}
\begin{split}
\frac{\partial \hat{V}}{\partial \mathbf{r}_i} = & \frac{\partial u_1(\mathbf{r}_i)}{\partial \mathbf{r}_i} + \frac{1}{2!}\sum_j\left( \frac{\partial u_2(\mathbf{r}_i,\mathbf{r}_j)}{\partial \mathbf{r}_i} + \frac{\partial u_2(\mathbf{r}_j,\mathbf{r}_i)}{\partial \mathbf{r}_i} \right) \\
& + \frac{1}{3!}\sum_{j,k}\left( \frac{\partial u_3(\mathbf{r}_i,\mathbf{r}_j,\mathbf{r}_k)}{\partial \mathbf{r}_i} + \frac{\partial u_3(\mathbf{r}_j,\mathbf{r}_i,\mathbf{r}_k)}{\partial \mathbf{r}_i} + \frac{\partial u_3(\mathbf{r}_j,\mathbf{r}_k,\mathbf{r}_i)}{\partial \mathbf{r}_i} \right) + \dots \\
= & \int d\mathbf{q}\,(\imath\,\mathbf{q}) \left( U_1^{(1)}(\mathbf{q}) + \frac{1}{2!}\sum_j \left( U_2^{(1)}(\mathbf{q},\mathbf{r}_j)+U_2^{(2)}(\mathbf{q},\mathbf{r}_j)\right) \right. \\
& \left. + \frac{1}{3!}\sum_{j,k}\left( U_3^{(1)}(\mathbf{q},\mathbf{r}_j,\mathbf{r}_k) + U_3^{(2)}(\mathbf{q},\mathbf{r}_j,\mathbf{r}_k) + U_3^{(3)}(\mathbf{q},\mathbf{r}_j,\mathbf{r}_k) \right) + \dots  \right)\,e^{\imath\,\mathbf{q}\cdot\mathbf{r}_i} \,,
\end{split}
\end{equation}
where we used Eq. (\ref{Fourierrep}). Since the many-body potentials are invariant for the permutation of their arguments, $U_n^{(1)}(\mathbf{q},\{\mathbf{r}_k\})=U_n^{(2)}(\mathbf{q},\{\mathbf{r}_k\})=\dots=U_n^{(n)}(\mathbf{q},\{\mathbf{r}_k\})$, and therefore Eq. (\ref{potentialexp}) reduces to:
\begin{equation}
\label{dVdri}
\begin{split}
\frac{\partial \hat{V}}{\partial \mathbf{r}_i} = \int d\mathbf{q}\,(\imath\,\mathbf{q})\,  & \left( U_1^{(1)}(\mathbf{q}) + \sum_j U_2^{(1)}(\mathbf{q},\mathbf{r}_j) \right. \\
& \left. + \frac{1}{2!}\sum_{j,k} U_3^{(1)}(\mathbf{q},\mathbf{r}_j,\mathbf{r}_k) + \dots  \right)\,e^{\imath\,\mathbf{q}\cdot\mathbf{r}_i} \,,
\end{split}
\end{equation}
Substituting Eq. (\ref{dVdri}) into Eq. (\ref{derG}), then rearranging the equation results in:
\begin{equation}
\label{momentk} 
\begin{split}
\partial_t \hat{\mathbf{G}}(\mathbf{k},t) & + (\imath\,\mathbf{k})\cdot\hat{\mathcal{K}}(\mathbf{k},t) \\
= & - \frac{1}{m} \int d\mathbf{q}\,(\imath\,\mathbf{q})\, \left( U_1^{(1)}(\mathbf{q}) + \sum_j U_2^{(1)}(\mathbf{q},\mathbf{r}_j) \right. \\
& \left. + \frac{1}{2!}\sum_{j,k} U_3^{(1)}(\mathbf{q},\mathbf{r}_j,\mathbf{r}_k) + \dots  \right)\hat{P}(\mathbf{k}-\mathbf{q},t) \,,
\end{split}
\end{equation}
where $\hat{\mathcal{K}}(\mathbf{k},t)=\frac{1}{(2\,\pi)^d}\sum_i \frac{\mathbf{p}_i \otimes \mathbf{p}_i}{m}\,e^{-\imath\,\mathbf{k}\cdot\mathbf{r}_i}$ is the Fourier transform of the microscopic kinetic stress $\hat{\mathbb{K}}(\mathbf{r},t) = \sum_i \frac{\mathbf{p}_i \otimes \mathbf{p}_i}{m}\,\delta[\mathbf{r}-\mathbf{r}_i(t)]$. The inverse Fourier transform of Eq. (\ref{momentk}) reads:
\begin{equation}
\label{momentApp}
\partial_t \hat{\mathbf{g}} + \nabla\cdot\hat{\mathbb{K}}= -\hat{\rho}\,\nabla \hat{\Phi} \\,
\end{equation}
where
\begin{equation}
\label{Phir}
\hat{\Phi}(\mathbf{r}) = \frac{1}{m}\,\left( u_1(\mathbf{r})+\sum_j u_2(\mathbf{r},\mathbf{r}_j) + \frac{1}{2!} \sum_{j,k} u_3(\mathbf{r},\mathbf{r}_j,\mathbf{r}_k) + \dots \right) \\.
\end{equation}
Similarly to the mass and momentum density, one can differentiate the Fourier transform of the microscopic kinetic energy density w.r.t. $t$, yielding:
\begin{equation}
\label{AppkinE0}
\begin{split}
\partial_t \hat{N}(\mathbf{k},t) = & \frac{1}{(2\,\pi)^d} \sum_i \left[ \frac{|\mathbf{p}_i|^2}{2m}\left(-\imath\,\mathbf{k}\cdot\frac{\mathbf{p}_i}{m}\right) + \frac{\mathbf{p}_i \cdot \dot{\mathbf{p}}_i}{m} \right] \,e^{-\imath\,\mathbf{k}\cdot\mathbf{r}_i} \\
= &  -\imath\,\mathbf{k}\cdot \hat{\mathbf{W}}(\mathbf{k},t)  - \frac{1}{(2\,\pi)^d} \sum_i \frac{\mathbf{p}_i}{m}\, \frac{\partial\hat{V}}{\partial \mathbf{r}_i} \,e^{-\imath\,\mathbf{k}\cdot\mathbf{r}_i} \\
= &  -\imath\,\mathbf{k}\cdot\hat{\mathbf{W}}(\mathbf{k},t) - \int d\mathbf{q}\,(\imath\,\mathbf{q})\,\Phi(\mathbf{q})\,\hat{\mathbf{G}}(\mathbf{k}-\mathbf{q},t) \,,
\end{split}
\end{equation}
where $\hat{\mathbf{W}}(\mathbf{k},t)=\frac{1}{(2\,\pi)^d} \sum_i \frac{|\mathbf{p}_i|^2}{2m}\,\frac{\mathbf{p}_i}{m} \,e^{-\imath\,\mathbf{k}\cdot\mathbf{r}_i}$ is the Fourier transform of the microscopic kinetic energy flux $\mathbf{w}(\mathbf{r},t)=\sum_i \frac{|\mathbf{p}_i|^2}{2m}\,\delta[\mathbf{r}-\mathbf{r}-i(t)]$, while $\Phi(\mathbf{q})$ is the Fourier transform of $\hat{\Phi}(\mathbf{r})$ defined by Eq. (\ref{Phir}). Finally, the inverse Fourier transform of Eq. (\ref{AppkinE0}) reads:
\begin{equation}
\label{thermalA}
\partial_t \hat{\kappa} + \nabla\cdot\hat{\mathbf{w}} = - \hat{\mathbf{g}} \cdot \nabla \hat{\Phi}\,.
\end{equation}

\section*{Appendix B. Coarse grained quantities}

The generic continuum equations read as:
\begin{eqnarray}
\label{CG1A}\partial_t \langle \hat{\rho} \rangle_t+ \nabla \cdot \langle \hat{\mathbf{g}} \rangle_t &=& 0 \,,\\
\label{CG2A}\partial_t \langle \hat{\mathbf{g}} \rangle_t+ \nabla \cdot \langle \hat{\mathbb{K}}  \rangle_t &=& -\langle \hat{\rho}\,\nabla\hat{\Phi} \rangle_t\,,\\
\label{CG3A}\partial_t \langle\hat{\kappa}\rangle_t + \nabla \cdot \langle\hat{\mathbf{w}}\rangle_t &=& -\langle\hat{\mathbf{g}}\cdot \nabla\hat{\Phi}\rangle_t \,,
\end{eqnarray}
where coarse-graining is defined as
\begin{equation}
\label{CGgen}
\langle . \rangle_t = \int_{\Omega} d\mathbf{\Gamma}\,(.)\,f(\mathbf{\Gamma},t) \,,
\end{equation}
and $f(\mathbf{\Gamma},t)$ is the full $N$-body p.d.f. (i.e., the solution of the Liouville equation). The single-particle quantities - i.e., all terms but the ones containing $\hat{\Phi}$) - in Eqns. (\ref{CG1A})-(\ref{CG3A}) will be evaluated for the one-particle density
\begin{equation}
\label{f1thermalA}
f^{(1)}(\mathbf{r}_i,\mathbf{p}_i,t) \approx  f^{(1)}_0(\mathbf{r}_i,\mathbf{p}_i,t)- f^{(1)}_v(\mathbf{r}_i,\mathbf{p}_i,t) - f^{(1)}_T(\mathbf{r}_i,\mathbf{p}_i,t) \,,
\end{equation} 
where
\begin{equation}
\label{LTEA}
f^{(1)}_0(\mathbf{r},\mathbf{p},t) = \frac{\rho(\mathbf{r},t)}{M}\,\frac{1}{(2\pi\,k_B \,T(\mathbf{r},t))^{d/2}}\,\exp\left({-\frac{|\mathbf{x}(\mathbf{p},\mathbf{r},t)|^2}{2\,k_B T(\mathbf{r},t)}}\right)
\end{equation} 
where $\mathbf{x}(\mathbf{p},\mathbf{r},t) = \mathbf{p} - m\,\mathbf{v}(\mathbf{r},t)$, while
\begin{eqnarray*}
f^{(1)}_v(\mathbf{r},\mathbf{p},t) &=& A(\lambda)\left( \mathbf{n}\cdot\mathbb{V}\cdot\mathbf{n} - \frac{1}{3}\,\nabla \cdot\mathbf{v} \right) \,,\\
f^{(1)}_T(\mathbf{r},\mathbf{p},t) &=& B(\lambda)\,(\mathbf{n}\cdot\mathbf\nabla T) \,,
\end{eqnarray*}
where $\lambda=|\mathbf{x}(\mathbf{p},\mathbf{r},t)|$ and $\mathbf{n}=\mathbf{x}(\mathbf{p},\mathbf{r},t)/\lambda$, and $\mathbb{V} = \frac{1}{2}\left[ (\nabla \otimes \mathbf{v}) + (\nabla \otimes \mathbf{v})^T \right]$. Since $f^{(1)}(\mathbf{r}_i,\mathbf{p}_i,t) $ is additive, one can start with the evaluation of the contributions emerging from LTE only. Accordingly:

\begin{itemize}
\item Mass density:
\begin{equation}
\begin{split}
\langle \hat{\rho}\rangle_t & = \int_\Omega d\mathbf{\Gamma}\, \sum_i m\,\delta(\mathbf{r}-\mathbf{r}_i) f(\mathbf{\Gamma},t) \\
&= m\sum_i \int d\mathbf{r}_i\,d\mathbf{p}_i \,\delta(\mathbf{r}-\mathbf{r}_i) \int \prod_{j \neq i}d\mathbf{r}_j\,d\mathbf{p}_j\,f(\mathbf{\Gamma},t) \\
&= m\sum_i \int d\mathbf{r}_i\,d\mathbf{p}_i \,\delta(\mathbf{r}-\mathbf{r}_i) f_0^{(1)}(\mathbf{r}_i,\mathbf{p}_i,t) \\
&= m\,N \int d\mathbf{p}\,f^{(1)}(\mathbf{r},\mathbf{p},t) = m\,N\, \frac{\rho}{M} = \rho(\mathbf{r},t) \,.
\end{split}
\end{equation}
\item Momentum density:
\begin{equation}
\begin{split}
\langle \hat{\mathbf{g}} \rangle_t &= \int_\Omega d\mathbf{\Gamma}\, \sum_i \mathbf{p}_i\,\delta(\mathbf{r}-\mathbf{r}_i)\,f(\mathbf{\Gamma},t) \\
&= N \int d\mathbf{p}\,\mathbf{p}\,f^{(1)}_0(\mathbf{r},\mathbf{p},t) \\
&= N \int d\mathbf{p}\,\left[\mathbf{x}(\mathbf{r},\mathbf{p},t)+m\,\mathbf{v}(\mathbf{r},t)\right]\,f^{(1)}_0(\mathbf{r},\mathbf{p},t) \\
&=  N\,m\,\mathbf{v}(\mathbf{r},t)\,\frac{\rho(\mathbf{r},t)}{M} = \rho(\mathbf{r},t)\,\mathbf{v}(\mathbf{r},t) \,.
\end{split}
\end{equation}
To simplify the notation, we will omit the arguments of $\mathbf{x}$ and the macroscopic fields in the intermediate formulae henceforth.
\item Kinetic energy density:
\begin{equation}
\begin{split}
\langle \hat{\kappa} \rangle_t &= \int_\Omega d\mathbf{\Gamma}\, \sum_i \frac{|\mathbf{p}_i|^2}{2\,m}\,\delta(\mathbf{r}-\mathbf{r}_i)\,f(\mathbf{\Gamma},t) \\
&= \frac{N}{2\,m} \int d\mathbf{p}\,|\mathbf{p}|^2\,f^{(1)}_0(\mathbf{r},\mathbf{p},t) \\
&= \frac{N}{2\,m} \int d\mathbf{p}\,|\mathbf{x}+m\,\mathbf{v}|^2\,f^{(1)}_0(\mathbf{r},\mathbf{p},t) \\
&=  \frac{N}{2\,m} \int d\mathbf{p}\,\left(|\mathbf{x}|^2+m^2\,|\mathbf{v}|^2+2\,m\,\mathbf{x}\cdot\mathbf{v}\right)\,f^{(1)}_0(\mathbf{r},\mathbf{p},t) \\
&=   \frac{N}{2\,m}\, \frac{\rho}{M}\, d\, m\,k_B T + \frac{N}{2\,m}\,m^2\,|\mathbf{v}|^2\,\frac{\rho}{M} \\
&=  \frac{d}{2}\,\frac{k_B T(\mathbf{r},t)}{m}\,\rho(\mathbf{r},t) + \frac{1}{2}\,\rho(\mathbf{r},t)\,|\mathbf{v}(\mathbf{r},t)|^2 \,,
\end{split}
\end{equation}
where the first term is the thermal energy density, the second the macroscopic kinetic energy density, and we used $\int d\mathbf{p}\,\mathbf{x}\,f_0^{(1)}(\mathbf{r},\mathbf{p},t)=\mathbf{0}$.
\item Kinetic stress tensor:
\begin{equation}
\begin{split}
\langle \hat{\mathbb{K}} \rangle_t &= \int_\Omega d\mathbf{\Gamma}\,\sum_i \frac{\mathbf{p}_i \otimes \mathbf{p}_i}{m}\,\delta(\mathbf{r}-\mathbf{r}_i)\,f(\mathbf{\Gamma},t) \\
&= \frac{N}{m} \int d\mathbf{p}\,\left( \mathbf{x}\otimes\mathbf{x} + m^2\,\mathbf{v}\otimes\mathbf{v}\right)\,f^{(1)}_0(\mathbf{r},\mathbf{p},t) \\
&= \frac{N}{m}\,\frac{\rho}{M}\,\left[(m\,k_B T)\,\mathbb{I} + m^2\,\mathbf{v}\otimes\mathbf{v} \right]\\
&= \frac{k_B T(\mathbf{r},t)}{m}\rho(\mathbf{r},t)\,\mathbb{I} + \rho(\mathbf{r},t)\,\mathbf{v}(\mathbf{r},t)\otimes\mathbf{v}(\mathbf{r},t) \,.
\end{split}
\end{equation}
\item Kinetic energy flux:
\begin{equation}
\begin{split}
\langle \hat{\mathbf{w}} \rangle_t & = \int_\Omega d\mathbf{\Gamma}\,\sum_i \frac{|\mathbf{p}_i|^2}{2\,m}\,\frac{\mathbf{p}_i}{m}\,\delta(\mathbf{r}-\mathbf{r}_i)\,f(\mathbf{\Gamma},t) \\
&= \frac{N}{2\,m^2} \int d\mathbf{p}\,\left[ 2\,(\mathbf{x} \otimes \mathbf{x})\cdot(m\,\mathbf{v}) + (|\mathbf{x}|^2 + m^2|\mathbf{v}|^2)\,\mathbf{v} \right]\,f^{(1)}_0(\mathbf{r},\mathbf{p},t) \\
&= \frac{N}{2\,m^2} \,\frac{\rho}{M}\left(2\, m^2\,k_B T\,\mathbf{v} + d\,m^2\,k_B T\,\mathbf{v}+ m^2 |\mathbf{v}|^2\mathbf{v} \right) \\
&= \left( \frac{d+2}{2}\,\frac{k_B T(\mathbf{r},t)}{m}\,\rho(\mathbf{r},t) + \frac{1}{2}\,\rho(\mathbf{r},t)|\mathbf{v}(\mathbf{r},t)|^2 \right)\mathbf{v}(\mathbf{r},t) \,,
\end{split}
\end{equation}
where we utilised the fact that the cubic moments of $\mathbf{x}$ vanish. 
\end{itemize}
To calculate the dissipative corrections to the coarse-grained quantities, first we investigate the moments of $\mathbf{p}$ w.r.t. $f_v(\mathbf{r},\mathbf{p},t)$ and $f_T(\mathbf{r},\mathbf{p},t)$. Since the coarse-graining process first replaces $\mathbf{r}_i$ by $\mathbf{r}$ (integral of Dirac-delta in a microscopic density), the only remaining integral appears in the form of
\begin{equation}
\label{Appgenmom}
\langle A(\mathbf{p}) \rangle_{\chi} \equiv \int d\mathbf{p}\,A(\mathbf{p})\,f^{(1)}_{\chi}(\mathbf{r},\mathbf{p},t) \, ,
\end{equation}  
Since both $f_{v}^{(1)}(\mathbf{r},\mathbf{p},t)$ and $f_{T}^{(1)}(\mathbf{r},\mathbf{p},t)$ depend on the length and direction of  $\mathbf{x}(\mathbf{r},\mathbf{p},t)=\mathbf{p}-m\,\mathbf{v}(\mathbf{r},t)$ separately, Eq. (\ref{Appgenmom}) can be easily evaluated by using the co-ordinate transformation $\mathbf{p} = \mathbf{x}+m\,\mathbf{v}(\mathbf{r},t)$ in $A(\mathbf{p})$, and then switching to spherical polar co-ordinates as $\mathbf{x}=\lambda \,\mathbf{n}$. Accordingly, the moments of $\mathbf{p}$ for $f_v(\mathbf{r},\mathbf{p},t)$ read (these can easily be checked by using Mathematica, for instance):
\begin{eqnarray}
\label{momentv1} \langle 1 \rangle_v &=& 0 \,, \\
\label{momentv2} \langle \mathbf{p} \rangle_v &=& \mathbf{0} \,,\\
\label{momentv3} \langle |\mathbf{p}|^2 \rangle_v  &=& 0 \,,\\
\label{momentv4} \langle \mathbf{p} \otimes \mathbf{p}\rangle_v &=& \tilde{\eta} \, \mathbb{D}\,, \\
\label{momentv5}  \langle |\mathbf{p}|^2\mathbf{p} \rangle_v &=& 2\,m\,\tilde{\eta}\,\mathbb{D}\cdot\mathbf{v} \,,
\end{eqnarray}
where $\tilde{\eta} = \frac{4\,\pi}{15} \int_0^\infty d\lambda\,\lambda^4 A(\lambda)$ is the reduced viscosity, and $\mathbb{D}=(\nabla\otimes\mathbf{v})+(\nabla\otimes\mathbf{v})^T-\frac{2}{3}(\nabla\cdot\mathbf{v})$ the viscous stress tensor. The moments of $\mathbf{p}$ over $f_T(\mathbf{r},\mathbf{p},t)$ can be calculated similarly, thus yielding:
\begin{eqnarray}
\label{momentT1}\langle 1 \rangle_T  &=& 0 \,, \\
\label{momentT2}\langle \mathbf{p} \rangle_T &=& \tilde{\omega}\,\nabla T \,, \\
\label{momentT3}\langle |\mathbf{p}|^2 \rangle_T  &=& 2\,m\,\tilde{\omega}\,\mathbf{v}\cdot\nabla T \,, \\
\label{momentT4}\langle \mathbf{p} \otimes \mathbf{p}_i \rangle_T &=& m\,\tilde{\omega}\,\left[ (\mathbf{v}\otimes\nabla T)+ (\nabla T\otimes\mathbf{v}) \right] \,, \\
\label{momentT5}\langle |\mathbf{p}|^2\mathbf{p} \rangle_T &=& \{\tilde{\kappa}+\tilde{\omega}\,m^2\,[|\mathbf{v}|^2\,\mathbb{I}+2\,(\mathbf{v}\otimes\mathbf{v})]\}\cdot\nabla T \,, \,
\end{eqnarray}
where $\tilde{\omega} = \frac{4\,\pi}{3}\int_0^\infty d\lambda\,\lambda^d B(\lambda)$ is a transport coefficient, and $\tilde{\kappa}=\frac{4\,\pi}{3}\int_0^\infty d\lambda\,\lambda^5 B(\lambda)$ the reduced heat diffusivity. The coarse-grained densities can be calculated by using Eqns. (\ref{momentv1})-(\ref{momentT5}) in Eq. (\ref{CGgen}). The calculations yield:
\begin{itemize}
\item Mass density:
\begin{equation}
\begin{split}
\langle \hat{\rho} \rangle_t &= \int_\Omega d\mathbf{\Gamma}\,\sum_i m\, \delta(\mathbf{r}-\mathbf{r}_i)\,f(\mathbf{\Gamma},t) \\
&= \rho - m\,N (\langle 1 \rangle_v + \langle 1 \rangle_T ) = \rho \,.
\end{split}
\end{equation}
\item Momentum density:
\begin{equation}
\begin{split}
\langle \hat{\mathbf{g}} \rangle_t &= \int_\Omega d\mathbf{\Gamma}\,\sum_i \mathbf{p}_i \, \delta(\mathbf{r}-\mathbf{r}_i)\,f(\mathbf{\Gamma},t) \\
&= \rho\,\mathbf{v} - N\,(\langle \mathbf{p}\rangle_v + \langle \mathbf{p}\rangle_T) = \rho\,\mathbf{v} - \omega\,\nabla T \,.
\end{split}
\end{equation}
\item Kinetic energy density:
\begin{equation}
\begin{split}
\langle \hat{\kappa} \rangle_t &= \int_\Omega d\mathbf{\Gamma}\,\sum_i \frac{|\mathbf{p}_i|^2}{2\,m} \, \delta(\mathbf{r}-\mathbf{r}_i)\,f(\mathbf{\Gamma},t) \\
&= q - \frac{N}{2\,m} \, (\langle |\mathbf{p}^2| \rangle_v + \langle |\mathbf{p}^2| \rangle_T )= q + \lambda - \omega\,\mathbf{v}\cdot\nabla T \,,
\end{split}
\end{equation}
where $q=\frac{d}{2}\,\frac{k_B T}{m}\,\rho$ and $\lambda = \frac{1}{2}\,\rho\,|\mathbf{v}|^2$ are the thermal and macroscopic kinetic energy density, respectively. 
\item Kinetic stress tensor:
\begin{equation}
\begin{split}
\langle \hat{\mathbb{K}} \rangle_t &= \int_\Omega d\mathbf{\Gamma}\,\sum_i \frac{\mathbf{p}_i \otimes \mathbf{p}_i}{m} \, \delta(\mathbf{r}-\mathbf{r}_i)\,f(\mathbf{\Gamma},t) \\
&= p_0\,\mathbb{I} + \rho\,\mathbf{v}\otimes\mathbf{v} - \frac{N}{m}\,(\langle \mathbf{p}\otimes\mathbf{p})_v + \langle \mathbf{p}\otimes\mathbf{p})_T )\\
&= p_0\,\mathbb{I} + \rho\,\mathbf{v}\otimes\mathbf{v}-\eta\,\mathbb{D} - \omega\,[(\mathbf{v} \otimes \nabla T)+(\nabla T \otimes \mathbf{v})] \,,
\end{split}
\end{equation}
where $p_0=\frac{k_B T}{m}\,\rho$ is the ideal gas pressure.
\item Kinetic energy flux:
\begin{equation}
\begin{split}
\langle \hat{\mathbf{w}} \rangle_t &= \int_\Omega d\mathbf{\Gamma}\,\sum_i \frac{|\mathbf{p}_i|^2}{2 m}\frac{\mathbf{p}_i}{m} \, \delta(\mathbf{r}-\mathbf{r}_i)\,f(\mathbf{\Gamma},t) \\
&= \left(q+p_0\right)\,\mathbf{v} - \frac{N}{2\,m^2} \,(\langle |\mathbf{p}|^2\mathbf{p} \rangle_v+\langle |\mathbf{p}|^2\mathbf{p} \rangle_T) \\
&=\left[(q+p_0)\,\mathbb{I} - \eta\,\mathbb{D}\right]\cdot\mathbf{v} -\left\{\chi+\omega\left[\frac{1}{2}|\mathbf{v}|^2\,\mathbb{I}+(\mathbf{v}\otimes\mathbf{v})\right]\right\}\cdot\nabla T \,,
\end{split}
\end{equation}
where $\eta=(N/m)\tilde{\eta}$ is the dynamic viscosity, $\chi=N/(2\,m^2)\tilde{\chi}$ the thermal conductivity, and $\omega=N\,\tilde{\omega}$ a transport coefficient. 
\end{itemize}
Before proceeding to the analysis of the hydrodynamic equations, we show that the terms containing $\omega$ in the coarse-grained quantities can be neglected in molecular fluids. Considering water at room temperature in 3 dimensions, for instance, and choosing $\tau \equiv \lambda\,\sqrt{m/(3\,k_B T_0)}$, $T_0 \equiv 300\,$K, and $\rho_0 \equiv 10^3\,$kg/m$^3$ as the unit of time, temperature, and density (respectively) results in the dimensionless viscosity $\hat{\eta} = 1.55\times 10^{-9}\,\textrm{m}/\lambda$, thermal conductivity $\hat{\chi} = 3.37 \times 10^{-9}\,\textrm{m}/\lambda$, and transport coefficient $\hat{\omega} = 4.2\times 10^{-55}\,\textrm{m}/\lambda$, where $\lambda$ is the unit of length. Here we used the particle mass $m=3\times10^{-26}\,$kg, the thermal conductivity $\chi=0.6\,$W/m$\,$K, and kinematic viscosity $\eta=0.001\,$Pas, and the unit of time was chosen so that the average of the dimensionless square velocity would be unity.) The result suggests that regardless of the length scale, the order of magnitude of $\hat{\omega}/\hat{\kappa}$ and $\hat{\omega}/\hat{\eta}$ is $10^{-46}$, and therefore any term containing $\omega$ in the coarse-grained equations is negligible. Summarising, the dominant parts of the coarse-grained quantities read as:
\begin{eqnarray}
\label{CGf1}\langle \hat{\rho} \rangle &=& \rho \,,\\
\label{CGf2}\langle \hat{\mathbf{g}} \rangle &=& \rho\,\mathbf{v} \,,\\
\label{CGf3}\langle \hat{\kappa} \rangle &=& q + \lambda \,,\\
\label{CGf4}\langle \hat{\mathbb{K}} \rangle &=& p_0\,\mathbb{I} + \rho\,\mathbf{v}\otimes\mathbf{v}-\eta\,\mathbb{D} \,,\\
\label{CGf5}\langle \hat{\mathbf{w}} \rangle &=&\left[(q+p_0)\,\mathbb{I} - \eta\,\mathbb{D}\right]\cdot\mathbf{v} - \chi\,\nabla T \,,
\end{eqnarray}
where $p_0 = \frac{k_B T}{m}\,\rho$, $q=\frac{d}{2}\,p_0$, $\lambda = \frac{1}{2}\,\rho\,|\mathbf{v}|^2$, and $\mathbb{D}=(\nabla\otimes\mathbf{v})+(\nabla\otimes\mathbf{v})^T-\frac{2}{3}(\nabla\cdot\mathbf{v})$.

\section*{Appendix C. Temperature equation and entropy production rate}

Using Eqns. (\ref{CGf1})-(\ref{CGf5}) and Eq. (\ref{Fcond}) in Eqns. (\ref{CG1A})-(\ref{CG3A}) results in the following  general coarse-grained equations:
\begin{eqnarray}
\label{Diss1A}\partial_t \rho + \nabla \cdot(\rho\,\mathbf{v}) &=& 0 \,,\\
\label{Diss2A}\partial_t(\rho\,\mathbf{v}) + \nabla\cdot(\rho\,\mathbf{v}\otimes\mathbf{v}) + \nabla p_0  - \nabla\cdot(\eta\,\mathbb{D})&=& - \langle\hat{\rho}\,\nabla\hat{\Phi}\rangle_t\,,\\
\label{Diss3A}\partial_t \kappa + \nabla\cdot[(\kappa+p_0)\mathbf{v}]- \nabla\cdot\,\mathbf{J} &=& -\mathbf{v}\,\cdot\langle\hat{\rho}\,\nabla\hat{\Phi}\rangle_t \,,
\end{eqnarray} 
where $\mathbf{J} = \eta\,\mathbb{D}\cdot\mathbf{v} + \chi\,\nabla T$. Using $\kappa = q+\lambda$ and Eq. (\ref{Diss2A}) in (\ref{Diss3A}) yields:
\begin{equation}
\label{energyexpand}
\begin{split}
[\partial_t q + \nabla\cdot(q\,\mathbf{v})] & + [\partial_t\lambda + \nabla\cdot(\lambda\,\mathbf{v})] \\
&= \mathbf{v}\cdot(\partial_t(\rho\,\mathbf{v}) + \nabla\cdot(\rho\,\mathbf{v}\otimes\mathbf{v})) -p_0\,(\nabla\cdot\mathbf{v}) + D \,,
\end{split}
\end{equation}
where $D = \eta\,\mathbb{D}:(\nabla \otimes \mathbf{v})+\nabla\cdot(\chi\,\nabla T)$.
Using $\lambda=\frac{1}{2}\,\rho\,|\mathbf{v}|^2$ together with the identities $\partial_t(\rho\,\mathbf{v})+\nabla\cdot(\rho\,\mathbf{v}\otimes\mathbf{v})=\rho\,[\partial_t \mathbf{v} + (\nabla\otimes\mathbf{v})\cdot\mathbf{v} ]$ (due to continuity) and $\nabla(|\mathbf{v}|^2) = 2(\nabla\otimes\mathbf{v})\cdot\mathbf{v}$, the second term on the left-hand side of Eq. (\ref{energyexpand}) can be further written as:
\begin{equation}
\label{thermalsimpA}
\begin{split}
\partial_t \lambda + \nabla\cdot(\lambda\,\mathbf{v}) & =\frac{|\mathbf{v}|^2}{2}\left( \partial_t \rho + \nabla\cdot(\rho\,\mathbf{v}) \right) + \frac{\rho}{2}\left( \partial_t(|\mathbf{v}|^2) + \mathbf{v}\cdot\nabla(|\mathbf{v}|^2) \right) \\
& = \rho\,\mathbf{v}\cdot\left( \partial_t \mathbf{v} + \frac{1}{2} \nabla(|\mathbf{v}|^2) \right) = \rho\,\mathbf{v}\cdot\left( \partial_t \mathbf{v} + (\nabla \otimes \mathbf{v})\cdot\mathbf{v} \right) \\
& =  \mathbf{v}\cdot \left( \partial_t(\rho\,\mathbf{v}) + \nabla\cdot(\rho\,\mathbf{v}\otimes\mathbf{v}) \right) \,,
\end{split}
\end{equation}
where we used the continuity equation in the last step. Using Eq. (\ref{thermalsimpA}) in Eq. (\ref{energyexpand}) yields: 
\begin{equation}
\label{middle}
[\partial_t q + \nabla\cdot(q\,\mathbf{v})] = -p_0\,(\nabla\cdot \mathbf{v}) + D \,.
\end{equation}
Using $q=\frac{d}{2}\,\frac{k_B T}{m}\,\rho$ in Eq. (\ref{middle}) results in:
\begin{equation}
\label{middle2}
\frac{d}{2}\,\frac{\rho\,k_B}{m} \left( \partial_t T + \mathbf{v}\cdot\nabla T\right) = -p_0\,(\nabla\cdot\mathbf{v}) + D \,,
\end{equation}
where we used continuity again. Re-arranging the above equation then yields: 
\begin{equation}
\label{temperatureA}
\frac{d}{2}\,\left( \partial_t \ln \left( \frac{T}{\tau} \right) + \mathbf{v}\cdot\nabla\ln\left(\frac{T}{\tau} \right) \right) = -\nabla\cdot \mathbf{v} + \frac{D}{p_0} \,.
\end{equation}
The entropy production rate can be calculated by using Eq. (\ref{temperatureA}) as follows. According to Eq. (\ref{AppSfull}), the entropy production rate reads:
\begin{equation}
\label{SfullA}
\dot{S} = \frac{k_B}{m}\,\frac{d}{dt} \int d\mathbf{r} \,\rho\, \left[1- \ln\left(\frac{\rho}{\delta} \right)+\frac{d}{2}\,\ln\left( \frac{T}{\tau}\right) \right] - \frac{d}{dt}\int d\mathbf{r}\,(\delta_T F_{exc}) \,.
\end{equation}
The first integral of Eq. (\ref{SfullA}) can be re-written by using
\begin{equation}
\label{AppSprod2}
\begin{split}
\frac{d}{dt}\int d\mathbf{r}\,\rho\left[ 1-\ln\left( \frac{\rho}{\delta}\right)\right] & = -\int d\mathbf{r}\,\ln\left( \frac{\rho}{\delta} \right) (\partial_t \rho) =  \int d\mathbf{r}\,\ln\left( \frac{\rho}{\delta} \right) \nabla\cdot(\rho\,\mathbf{v}) \\ &= \int d\mathbf{r}\,\nabla\cdot\left( \ln\left(\frac{\rho}{\delta}\right)\rho\,\mathbf{v} \right) - \int d\mathbf{r}\,\rho\,\mathbf{v}\cdot\nabla \ln\left(\frac{\rho}{\delta}\right)\\
& =  \oint d\mathbf{A}\cdot\ln\left(\frac{\rho}{\delta}\right)\rho\,\mathbf{v} - \int d\mathbf{r}\,\mathbf{v}\cdot\nabla\rho \,,\\
& = - \oint d\mathbf{A}\cdot \rho\,\mathbf{v} + \int d\mathbf{r}\,\rho\,\nabla\cdot\mathbf{v} = \int d\mathbf{r}\,\rho\,\nabla\cdot\mathbf{v} \,,
\end{split}
\end{equation}
where we utilised continuity, then Gauss' integral theorem, and also utilised that $\mathbf{v}\cdot d\mathbf{A}=0$ on the boundary of the volume (isolated system). The divergence of the velocity field can be expressed from Eq. (\ref{temperatureA}):
\begin{equation}
\label{Appident}
\nabla\cdot\mathbf{v} = \frac{D}{p_0} - \frac{d}{2}\left[ \partial_t \ln\left(\frac{T}{\tau}\right) + \mathbf{v}\cdot\nabla \ln\left(\frac{T}{\tau} \right) \right] \,.
\end{equation}
Substituting Eq. (\ref{Appident}) into Eq. (\ref{AppSprod2}), using the identities
\begin{eqnarray}
\rho\,\partial_t \ln\left(\frac{T}{\tau}\right) &=& \frac{d}{dt}\left[ \rho\,\ln\left( \frac{T}{\tau} \right) \right] - (\partial_t \rho)\,\ln\left( \frac{T}{\tau} \right) \,, \\
\rho\,\mathbf{v}\cdot\nabla \ln\left(\frac{T}{\tau}\right) &=& \nabla\cdot\left[ \rho\,\mathbf{v}\,\ln\left( \frac{T}{\tau} \right) \right] - \nabla\cdot(\rho\,\mathbf{v})\,\ln\left( \frac{T}{\tau} \right) \,,
\end{eqnarray}
then recalling Gauss' theorem again gives:
\begin{equation}
\label{firstintegral}
\frac{d}{dt} \int d\mathbf{r}\,\rho\left[ 1-\ln\left( \frac{\rho}{\delta}\right)\right] = \int d\mathbf{r}\,\left(\frac{\rho}{p_0}D\right) - \frac{d}{2}\frac{d}{dt} \int d\mathbf{r}\,\rho\,\ln\left( \frac{T}{\tau} \right) \,.
\end{equation}
Finally, using Eq. (\ref{firstintegral}) in Eq. (\ref{SfullA}), and re-writing the terms by following Landau \& Lifshitz \cite{Landau1987Fluid} result in:
\begin{equation}
\label{SprodA}
\dot{S} = \int d\mathbf{r}\,\left( \kappa\,\left| \frac{\nabla T}{T} \right|^2 + \frac{\eta}{2\,T} (\Sigma\mathbb{D})^2\right) - \frac{d}{dt}\int d\mathbf{r}\,(\delta_T F_{exc} ) \,.
\end{equation} 

\nocite{*}
\bibliography{papers}

\end{document}